\documentclass[aps,prl,showpacs,twocolumn]{revtex4}
\usepackage{graphicx,epsfig}
\usepackage{dcolumn}
\usepackage{amsmath}
\usepackage{amssymb}
\usepackage{epstopdf}
\usepackage{amsmath,verbatim}
\usepackage{bm}

\def\kv{{\bf k}}
\def\qv{{\bf q}}

\def\Gcal{{\bf {\cal G}}}

\def\nn{\nonumber}
\begin{document}
\title{Theory of time-resolved spectral function in high-temperature superconductors
with bosonic modes}
\author{Jianmin Tao}
\author{Jian-Xin Zhu}
\affiliation{Theoretical Division \& CNLS,
Los Alamos National Laboratory, Los Alamos, New Mexico 87545}
\date{\today}
\begin{abstract}
We develop a three-temperature model to simulate the time dependence of electron 
and phonon temperatures in high-temperature superconductors displaying strong
anistropic electron-phonon coupling. 
This model not only takes the tight-binding band structure into account, 
but also is valid in superconducting state. Based on this model, we calculate the
time-resolved spectral function via the double-time Green's functions. We find
that the dip-hump structure evolves with the time delay. More interestingly, new
phononic structures are obtained when the phonons are excited by a laser field.
This signature may serve as a direct evidence for electron-vibration mode
coupling.
\end{abstract}
\maketitle

The discovery of high-temperature superconductors (HTSC) has raised an important
issue on the mechanism leading to the formation of Cooper pairs, which is still under 
debate. To address this issue, a number of spectroscopy 
 techniques~\cite{ADamascelli03,JLee06,FCarbone08}
have been used to study the role and nature of bosonic modes, 
to which electrons are strongly coupled. 
In particular, due to technological advances 
and the improved sample quality, angle-resolved photoemission spectroscopy
(ARPES)~\cite{ADamascelli03} 
has been used to probe details of the energy and momentum 
structure of single-particle excitations via the measurement of 
photoemission intensity. However, different 
interpretations of the same data may lead to completely different 
mechanisms~\cite{MRNorman01,AWSandvik04}. 
For example, the dip-hump 
structure observed in 
ARPES~\cite{HDing96,JCCampuzano99,AKaminski01,ALanzara01,PDJohnson01,
XJZhou03,TSato03,TCuk04} could be interpreted~\cite{ZXShen97} as naturnally 
occurring in the interacting system but having little effect 
on the superconducting pairing mechanism. 
It could also be interpreted~\cite{ZXShen02} in terms of phonon modes that could drive
$d$-wave pairing. 
Recent theoretical analysis~\cite{TPDevereaux04} on the spectral function of
thermally excited electrons in the cuprates has shown that the 
out-of-plane and out-of-phase buckling mode strongly couples to 
the electronic states near the anti-nodal $M$ points in the Brillouin zone, while the in-plane breathing 
mode couples strongly to the electronic states near the $d$-wave nodal points. The signature 
of the anistropic electron-phonon (el-ph) coupling seems to get enhanced in the
superconducting state, suggesting the significant role of the 
el-ph interaction in the superconducting mechanism. 

The information extracted from the conventional ARPES is limited. 
Time-resolved ARPES offers the capability to simultaneously capture 
the single-particle (frequency domain) and collective (time domain) 
information, thus making it possible to directly probe the link 
between the collective modes and single-particle states. 
In this setting, either electrons or lattice vibrational modes 
can be selectively excited with an ultrafast laser pulse. Recent 
applications of this technique include the studies of transient 
electronic structure in Mott insulators~\cite{LPerfetti06,FSchmitt08} and high-$T_c$ 
cuprates~\cite{LPerfetti07} with optical pump. Furthermore, direct pumping of 
vibrational mode has also been realized in manganites~\cite{MRini07}, though not 
yet  in the cuprates. 
Motivated by the thrust of this experimental technique, in this Letter, 
we aim to provide a theoretical underpinning of transient electronic structure 
for HTSC, in a hope to better understand the nature of bosonic modes in these systems.
As such, the time evolution of the el-ph coupling
is investigated for both normal and superconducting states
with the time-resolved spectral function. Our theory consists of two parts.
First, we develop a three-temperature model
to simulate the time dependence of the
electron and phonon temperatures. 
Then, based on the three-temperature model,
we calculate the time-resolved spectral function.
Our results show a kink-structure in the time dependence of the 
electronic temperature at the superconducting transition temperature. 
Accordingly, we find from the spectral density that 
the energy position of the phonon mode is offset by a time-dependent gap function.  
More interestingly, new signatures of the el-ph coupling, which
are absent when electrons are excited, can be observed 
in the case of selective excitation of phonons.
This signature may serve as a direct evidence for the 
electron-vibration mode coupling as opposed to the coupling between electrons and 
spin fluctuations.
 
{\it Three-temperature model}: 
Consider a two-dimensional superconductor exposed to a laser field.
The model Hamiltonian  for a vibrational mode $\nu$ can be written as
\begin{eqnarray}\label{hamiltonian}
H &=& \sum_{\kv\sigma}\xi_{\kv}c^{\dagger}_{\kv\sigma}
c_{\kv\sigma}+
\sum_{\kv}(\Delta_{\kv} c_{\kv\uparrow}^{\dagger}
 c_{-\kv\downarrow}^{\dagger}+{\rm h.c.}) 
+
\sum_{\qv}\hbar\Omega_{\nu\qv} \nn \\
&\times&
\bigg(b_{\nu\qv}^{\dagger} b_{\nu\qv}+\frac{1}{2}\bigg)
+
\frac{1}{\sqrt{N_L}}\sum_{\kv\qv\sigma}g_\nu(\kv,\qv)
c^{\dagger}_{\kv+\qv,\sigma} c_{\kv\sigma} A_{\nu\qv} \nn \\
&+& H_{\rm field}(\tau),
\end{eqnarray}
where $c^{\dagger}_{\kv\sigma}$ ($b_{\nu\qv}^{\dagger}$) and $c_{\kv\sigma}$ 
($b_{\nu\qv}$) are the creation and annihilation operators for an electron 
with momentum $\mathbf{k}$ and
spin $\sigma$ (phonon with momentum $\mathbf{q}$ and vibrational mode $\nu$), 
$ A_{\nu\qv} =  b_{-\nu\qv}^{\dagger}+ b_{\nu\qv}$, the quantity 
$\xi_{\mathbf{k}}$  is the normal-state energy dispersion,
 $\mu$ the chemical potential, 
$\Delta_\kv$  the gap function,  and $g_\nu$ the coupling 
matrix.

By performing the Bogoliubov-de Gennes transformation~\cite{PGdeGennes66}, 
$c_{\kv\uparrow} = 
u_\kv\alpha_\kv-v_\kv\beta_\kv^{\dagger}$ and
$c_{-\kv\downarrow} = u_\kv\beta_\kv+v_\kv\alpha_\kv^{\dagger}$, 
we obtain
$E_e =\sum_\kv E_\kv
(\langle \alpha_\kv^\dagger \alpha_\kv \rangle -
\langle \beta_\kv \beta_\kv^\dagger \rangle) $,
where $E_\kv = \sqrt{\xi_\kv^2+\Delta_\kv^2}$. The time 
evolution of $\langle \alpha_\kv^\dagger \alpha_\kv \rangle$
and $\langle \beta_\kv \beta_\kv^\dagger \rangle$ is calculated
using the Heisenberg equations-of-motion approach. 
They are found to be
\begin{eqnarray}\label{scattering}
\frac{\partial \langle \alpha_\kv^\dagger \alpha_\kv \rangle}{\partial \tau} &=&
\frac{2\pi}{N_L}\sum_\qv g_\nu^2(u_\kv u_{\kv-\qv}-v_\kv v_{\kv-\qv})^2
(\delta_2 e^{\beta_{\rm e}\Omega_0}-\delta_1)
\nn \\
&\times&
\bigg[e^{(\beta_{ph}-\beta_e)\Omega_0}-1\bigg]
(1-f_{\mathbf{k}})f_{\mathbf{k}-\mathbf{q}} N_{\Omega_0}\;, 
\end{eqnarray}
and $\partial \langle \beta_\kv \beta_\kv^\dagger \rangle/\partial \tau =
-\partial \langle \alpha_\kv^\dagger \alpha_\kv \rangle/\partial \tau$,
where 
$f_{\mathbf{k}} = f(E_\kv)=1/(e^{\beta_e E_\kv}+1)$, $N_{\Omega_0}=
N(\Omega_0)=1/(e^{\beta_{ph} \Omega_0}-1)$,
$\delta_1 = \delta (E_{\kv-\qv}-E_\kv+\Omega_0)$, and
$\delta_2 = \delta (E_{\kv-\qv}-E_\kv-\Omega_0)$. Here we have specifically set  $\Omega_\nu = \Omega_0$.
Differentiation of both sides of the expression for $E_e$ and substitution of 
Eq.~(\ref{scattering})  leads to 
the rate of energy exchange 
\begin{eqnarray}\label{rate}
\frac{\partial E_e}{\partial \tau} &=&
\frac{4\pi}{N_L}\sum_\qv g_\nu^2(u_\kv u_{\kv-\qv}-v_\kv v_{\kv-\qv})^2
\delta (E_{\kv-\qv}-E_\kv-\Omega_0) \nn \\
&\times&
\Omega_0 \bigg[e^{(\beta_{ph}-\beta_e)\Omega_0}-1\bigg]
f_{\mathbf{k}}(1-f_{\mathbf{k}-\mathbf{q}})N_{\Omega_0}\;.
\end{eqnarray}
Recent experiment~\cite{LPerfetti07} shows that there exists cold lattice, which is 
negligibly coupled to the electrons but dissipates the energy of hot phonons via anharmonic 
cooling. Considering this observation, we write a set of rate equations for  the electron,
hot phonon, and lattice temperatures as
\begin{eqnarray}\label{three-Te}
\frac{\partial T_e}{\partial \tau} &=& \frac{1}{C_e}\frac{\partial E_e}{\partial \tau}
+\frac{P_e}{C_e}, \\
\label{three-Tph}
\frac{\partial T_{ph}}{\partial \tau} &=& - \frac{1}{C_{ph}}
\frac{\partial E_e}{\partial \tau} + \frac{P_{ph}}{C_{ph}} - 
\frac{T_{ph}-T_l}{\tau_\beta}, \\
\label{three-Tl}
\frac{\partial T_l}{\partial \tau} &=& 
\bigg(\frac{C_{ph}}{C_l}\bigg)\frac{T_{ph}-T_l}{\tau_\beta},
\end{eqnarray}
where $P_e$ is the power for pumping electrons and $P_{ph}$ the power 
for pumping hot phonons. 
The specific heat of electrons can be calculated from 
the Boltzmann entropy
$S_e=-2k_B\sum_\kv\{[1-f(E_\kv)]{\rm ln}[1-f(E_\kv)]+f(E_\kv) {\rm ln} f(E_\kv)\}$ 
by $C_e= T_e\partial S_e/\partial T_e$, while 
the specific heat of hot phonons for {\it one}-vibrational mode can be 
calculated from the simple relationship 
$C_{ph}=\hbar \Omega \partial N(\Omega)/\partial T_{ph}\vert_{\Omega = \Omega_0}$.
The results are given by 
\begin{eqnarray}
\label{eheat}
C_e &=& 
\beta k_B\sum_\kv\bigg(-\frac{\partial f(E_\kv)}{\partial E_\kv}\bigg)\bigg(2E_\kv^2+
\beta \Delta_k \frac{\partial \Delta_k}{\partial \beta}\bigg),~~~~ \\
\label{phheat}
C_{ph} &=& \frac{k_B}{4}\bigg(\frac{\hbar \Omega_0}{k_B T_{ph}}\bigg)^2
\bigg[{\rm coth}^2\bigg(\frac{\hbar \Omega_0}{2 k_B T_{ph}}\bigg)-1\bigg],
\end{eqnarray}
respectively, with $\coth x \equiv (e^x+e^{-x})/(e^x-e^{-x})$.
Equations~(\ref{rate})-(\ref{phheat}) constitute our three-temperature model.
The original version of this model
was phenomenologically proposed~\cite{LPerfetti07} as an extension of the 
two-temperature model~\cite{PBAllen87} for the normal state. The present model has the following advantages:
(i) it incorporates the detailed band structure, (ii) it is valid for {\it both}  
normal and superconducting states, and (iii) it includes the anistropic
effect on the el-ph coupling. 

\begin{figure}
\centerline{\psfig{figure=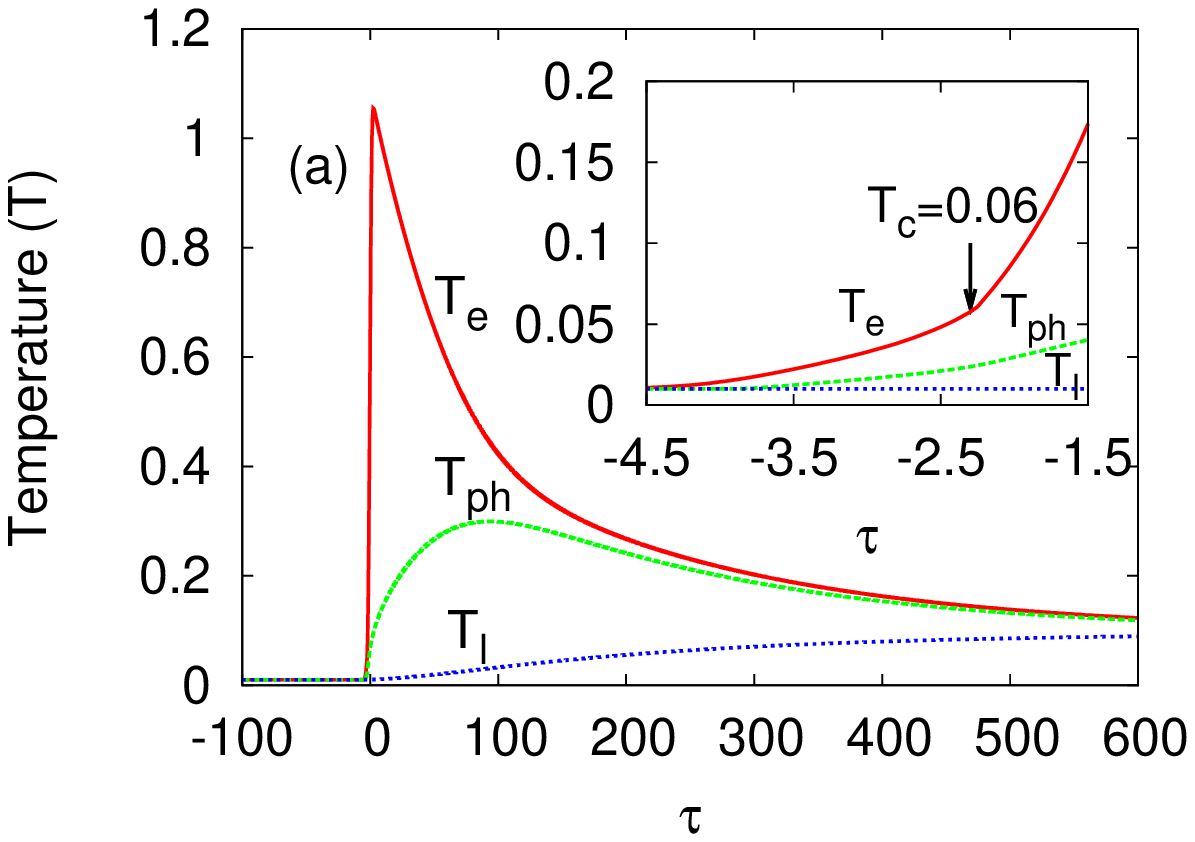,width=4.0cm,angle=0}
\psfig{figure=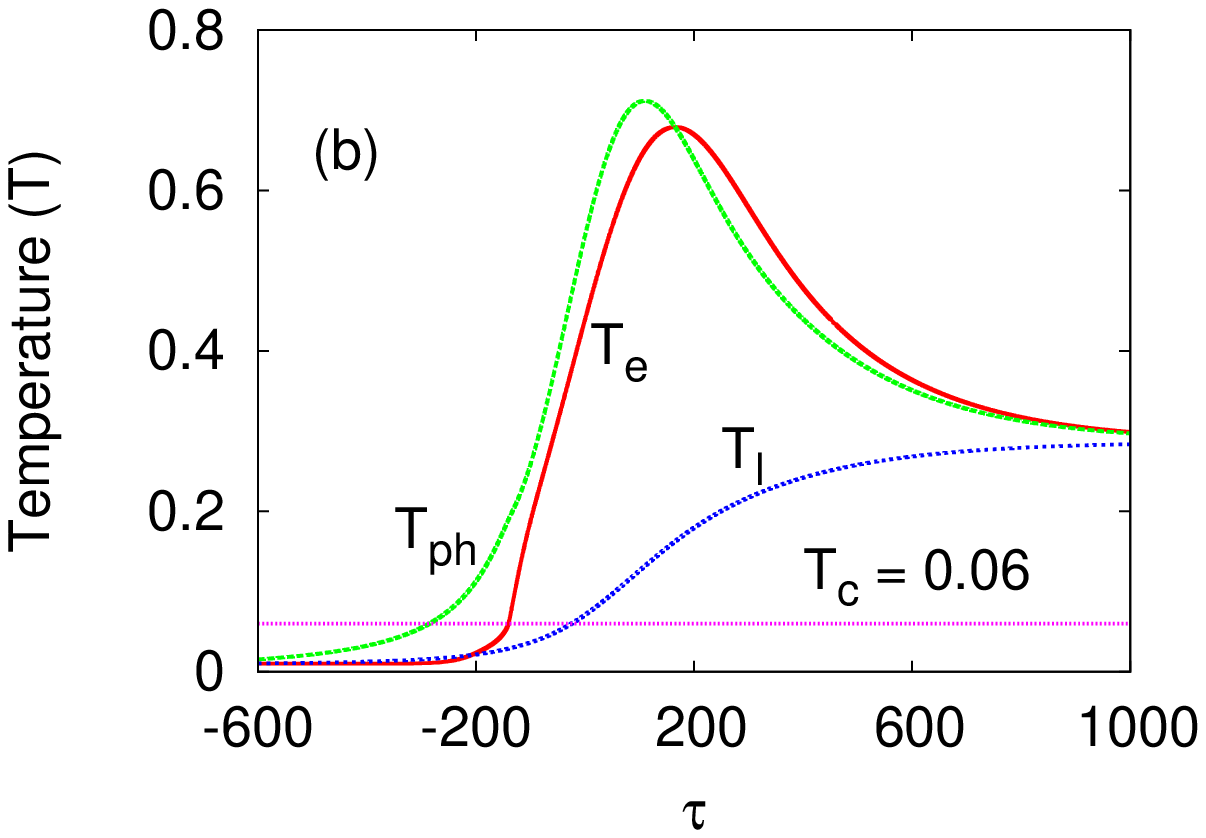,width=4.0cm,angle=0}}
\caption{(Color) Time evolution of electron ($T_e$), phonon ($T_{ph}$), 
and lattice ($T_l$) temperatures for selectively exciting electrons
(a) and phonons (b) with a laser pulse. }
\label{temp}
\end{figure}

Now we apply our three-temperature model to simulate the time 
evolution of the electron, hot phonon, and lattice temperatures 
for a $d$-wave superconductor.
We use a five-parameter tight-binding model~\cite{MRNorman95}
 to describe the energy dispersion: $\xi_{\mathbf{k}} = -2t_{1} (\cos k_x + \cos k_y)  - 4t_2 \cos k_x \cos k_y - 2t_3 (\cos 2k_x + \cos 2k_y) - 4t_4 (\cos 2k_x \cos k_y + \cos k_x \cos 2k_y) -4t_5 \cos 2k_x \cos 2k_y -\mu$, where 
$t_1=1$, $t_2=-0.2749$, $t_3=0.0872$, $t_4=0.0938$, $t_5=-0.0857$, and $\mu=-0.8772$. 
Unless specified explicitly,  the energy is measured in units of $t_1$ 
and the time is measured in units of $\hbar/t_1$ hereafter. 
(For $t_1=150$ meV, it corresponds to 1740 Kelvin in temperature and 
$\hbar/t_1$ to $4.4$ femtosecond in time.)
 The $d$-wave gap function has the form
$\Delta_k = \Delta_0(T_e)({\rm cos}k_x-{\rm cos}k_y)/2$.
The temperature-dependent part is given by~\cite{FGross86} 
$\Delta_0(T_e) = \Delta_{00} {\rm tanh}\{(\pi/z)\sqrt{ar(T_c/T_e-1)}\}$,
where $z=\Delta_{00}/(k_BT_c)$ and
$\tanh x = 1/\coth x$. 
In our calculations, we
set $\Delta_{00} = 0.2$, the critical temperature $T_c = 0.06$,
the specific heat jump at $T_c$ is $r=\Delta C_e/C_e \sim 1.43$,
and $a=2/3$.
In this work, we focus on
the buckling phonon mode, for 
which~\cite{TPDevereaux04,JXZhu06a,JXZhu06b}
$g_{B_{1g}} = g_0\{[{\rm cos}^2(q_x/2)+{\rm cos}^2(q_y/2)]/2\}^{-1/2}
\{\phi_x(\kv)\phi_x(\kv+\qv) 
{\rm cos}(q_y/2)-\phi_y(\kv)\phi_y(\kv+\qv)
{\rm cos}(q_x/2)\}$
with $\phi_x=(i/N_\kv)[\xi_{\kv}t_{x,\kv}-t_{xy,\kv}t_{y,\kv}]$,
$\phi_y=(i/N_k)[\xi_{\kv}t_{y,\kv}-t_{xy,\kv}t_{x,\kv}]$,
$N_\kv = [(\xi_{\kv}^2-t_{xy,\kv}^2)^2+(\xi_{\kv}t_{x,\kv}-t_{xy,\kv}t_{y,\kv})^2
+(\xi_{\kv}t_{y,\kv}-t_{xy,\kv}t_{x,\kv})^2]^{1/2}$,
$t_{\alpha,\kv}=-2t_1 {\rm sin}(k_{\alpha}/2)$, and
$t_{xy,\kv} = -4t_2 {\rm sin}(k_x/2){\rm sin}(k_y/2)$. 
To be consistent with experiment~\cite{LPerfetti07,TPDevereaux04}, 
we set $g_0 = 0.4$ and 
 the mode frequency $\Omega_0 = 0.3$.
The relaxation time $\tau_\beta = 200$. The
pump power is a Gaussian pulse of  
$P=P_0 e^{-\tau^2/(2\sigma^2)}$,
for which, the corresponding FWHM (full width at half maximum) is 2.35 $\sigma$. 
To pump electrons, we set $P_0 = 0.15$ and $\sigma = 1$.
Considering that the energy scale of phonons is much smaller than that of electrons,
we set $P_0 = 0.006$ and $\sigma = 100$ for pumping phonons.
The ratio $C_{ph}/C_l$ in Eq.~(\ref{three-Tl}) is set to be 0.2. 

Figure~\ref{temp} displays the temporal evolution of three respective 
temperatures for pumping electrons (a) and hot phonons (b).  
Before pumping, all three types of degrees of freedom (DoF) are in equilibrium, 
which is set at $T_e=T_{ph}=T_l=0.01$. As shown in Fig.~\ref{temp}(a), 
when  the pump pulse is absorbed by the electrons, the electronic temperature 
rises steeply around $\tau = 0$, and reaches the maximum after a small 
time delay. It then begins to drop at a time scale determined by the el-ph 
coupling strength, followed by a slower relaxation.  Interestngly,  
we also observe a small kink in $T_e$ at $T_e = T_c$, which entirely arises  
from the breakup of the Cooper pairs, resulting in the dramatic change
in the rise rate of $T_e$. This kink was not captured 
in the early model~\cite{LPerfetti07}. Simultaneously, the
hot phononic temperature $T_{ph}$ rises smoothly via the energy exchange 
with electrons and then drops by exchanging energy with
the cold lattice, causing the slight increase of $T_l$. 
When directly pumping
phonons, the time dependence of the temperatures is similarly observed,
including the kink in $T_e$, as shown in Fig.~\ref{temp}(b). 
Due to the large pumping width, the kink is more easily visible
in this case.   

{\it Time-resolved spectral function}: The time-resolved spectral
function is defined as
$A(\kv,\omega)\equiv -\frac{2}{\pi}~ {\rm Im} \Gcal_{11}(\kv,\omega)$.
Here $\Gcal_{11}$ is the one-one component of the retarded Green's 
function ${\hat \Gcal}(\kv,\omega)$, which is
related to the self-energy by
${\hat \Gcal}^{-1}(\kv,\omega) = {\hat \Gcal}_0^{-1}(\kv,\omega) - 
{\hat\Sigma}(\kv,\omega)$,
with ${\hat \Gcal}_0^{-1}(\kv,\omega) = \omega {\hat\sigma}_0
-\Delta_\kv {\hat\sigma}_1 - \xi_\kv {\hat\sigma}_3$.
${\hat\sigma}_{0,1,2,3}$ are the unit and Pauli matrices.
As is known~\cite{TPDevereaux04}, in the equilibrium state with $T_e=T_{ph}$, 
the self-energy can be evaluated more conveniently within the imaginary-time 
Green's function approach, in which  
a key step is to convert the Bose-Einstein distribution to the Fermi distribution,
$n_B(i\omega_n\pm E_{\kv-\qv}) = -n_F(\pm E_{\kv-\qv})$ (with $\omega_{n}=(2n+1)\pi T$,
$T=T_e=T_{ph}$). 
For the current situation, the temperatures of electrons and hot phonons 
are no longer tied to each other and the above conversions are
not valid any more. To avoid this restriction, 
here we apply the double-time Green's function approach~\cite{DNZubarev60} to 
calculate ${\hat\Sigma}(\kv,\omega)$. The result is 
\begin{eqnarray}\label{selfe}
{\hat\Sigma}(\kv,\omega) &=& \frac{1}{N_L}\sum_{\qv}\vert g_\nu(\mathbf{k},-\mathbf{q}) \vert^2 
\nonumber \\
&&\{  [(\omega-\Omega_0)\Phi_{1} -
(\omega+\Omega_0)\Phi_{2} +
\Omega_0 
(\Phi_{3} + \Phi_{4})]{\hat\sigma}_0 \nn \\
&& +
[2 E_{\kv-\qv}(\Phi_{1}-\Phi_{3})
+
2\Omega_0(\Phi_{3}-\Phi_{4})]{\hat\sigma}_1 
\nn \\
&&+  
[\xi_\kv(\Phi_{1}-\Phi_{2})+(\Omega_0\xi_\kv/E_\kv)(\Phi_{3}-\Phi_{4})] 
{\hat\sigma}_3\},
\end{eqnarray}
where 
$\Phi_{1}=N(\Omega_0)/[(\omega-\Omega_0+E_{\kv-\qv})(\omega-\Omega_0-E_{\kv-\qv})]$,
$\Phi_{2}=N(-\Omega_0)/[(\omega+\Omega_0+E_{\kv-\qv})(\omega+\Omega_0-E_{\kv-\qv})]$,
$\Phi_{3}=f(-E_{\kv-\qv})/[(\omega+\Omega_0-E_{\kv-\qv})(\omega-\Omega_0-E_{\kv-\qv})]$,
and
$\Phi_{4}=f(E_{\kv-\qv})/[(\omega+\Omega_0+E_{\kv-\qv})(\omega-\Omega_0+E_{\kv-\qv})]$.

\begin{figure}
\centerline{\psfig{figure=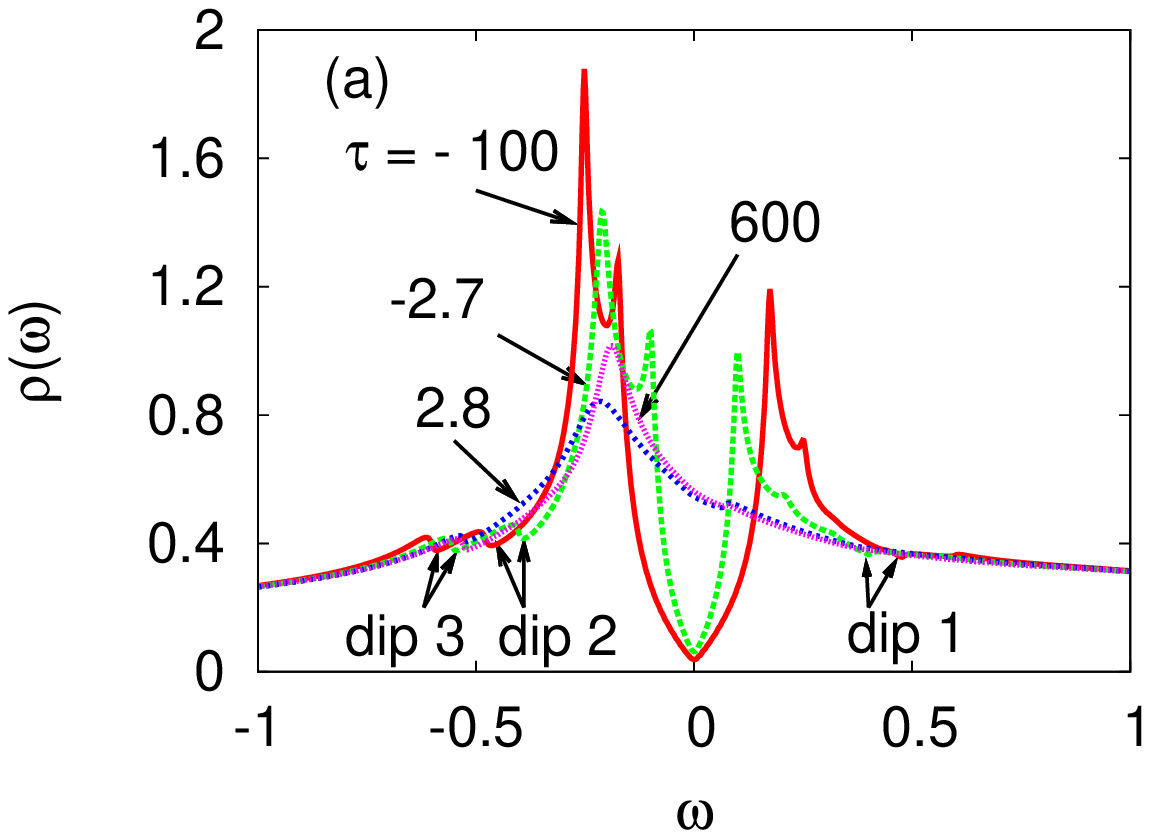,width=4.0cm,angle=0}
\psfig{figure=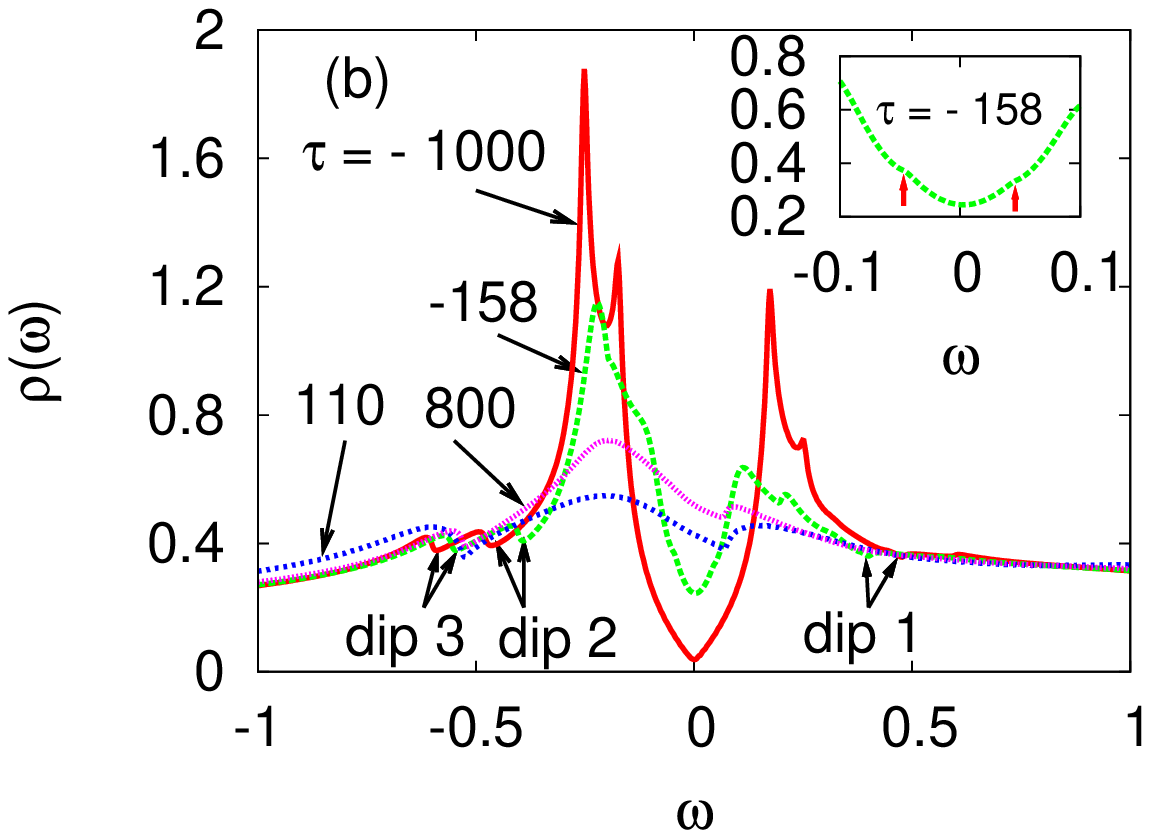,width=4.0cm,angle=0}}
\caption{(Color) Time evolution of the density of states for selectively
exciting electrons (a) and phonons (b), respectively.} 
\label{dosf}
\end{figure}

For simplicity, let us first look at the density of states (DOS), 
which is defined by
$\rho(\omega) = \frac{1}{N_L}\sum_\kv A(\kv,\omega)$.
For direct pumping of electrons,
we calculate the DOS at time sequences $\tau = -100$,
$-2.7$, $2.8$, and $600$, respectively. The results are plotted 
in Fig.~\ref{dosf} (a). At the initial time $\tau = -50$ where $T_e = 0.01$
and the system is at superconducting state [$\Delta_0(T_e) = 0.19$], 
we observe two signatures of the el-ph coupling located 
at $\pm (\Omega_0 + \Delta_{0}(T_e))$,
as shown in Fig.~\ref{dosf} (dip 1 and dip 2). They are
symmetric with respect to $\omega=0$, but dip 2 is much stronger. It is related 
to the fact that the normal-state van Hove singularity is located below the 
Fermi energy and as such the coherent peak at $-\Delta_0(T_e)$ has stronger 
intensity than that at $\Delta_0(T_e)$. 
In addition,  dip 3 occurs near the characteristic energy $-(E_M + \Omega_0)$, 
where $E_M$ is the quasiparticle energy at $\mathbf{k}=(\pi,0)$ or equivalent 
wavevector points in Brillouin zone.  It arises from the van Hove singularity 
at $\mathbf{k}=(\pi,0)$. The signature at the energy $E_M + \Omega_0$ is much 
weaker also because of the van Hove singularity location in the normal state.  
At $\tau = -2.7$ where $T_e = 0.041$ and $\Delta_0(T_e) \approx 0.1$, 
similar behaviors are observed. However, at $\tau = 2.8$, $T_e=1.05$ 
and thus only the signature of the normal-state el-ph coupling
(dip 3) can be observed. At $\tau = 600$, a similar structure of the normal-state
el-ph coupling is observed.

\begin{figure}
\centerline{\psfig{figure=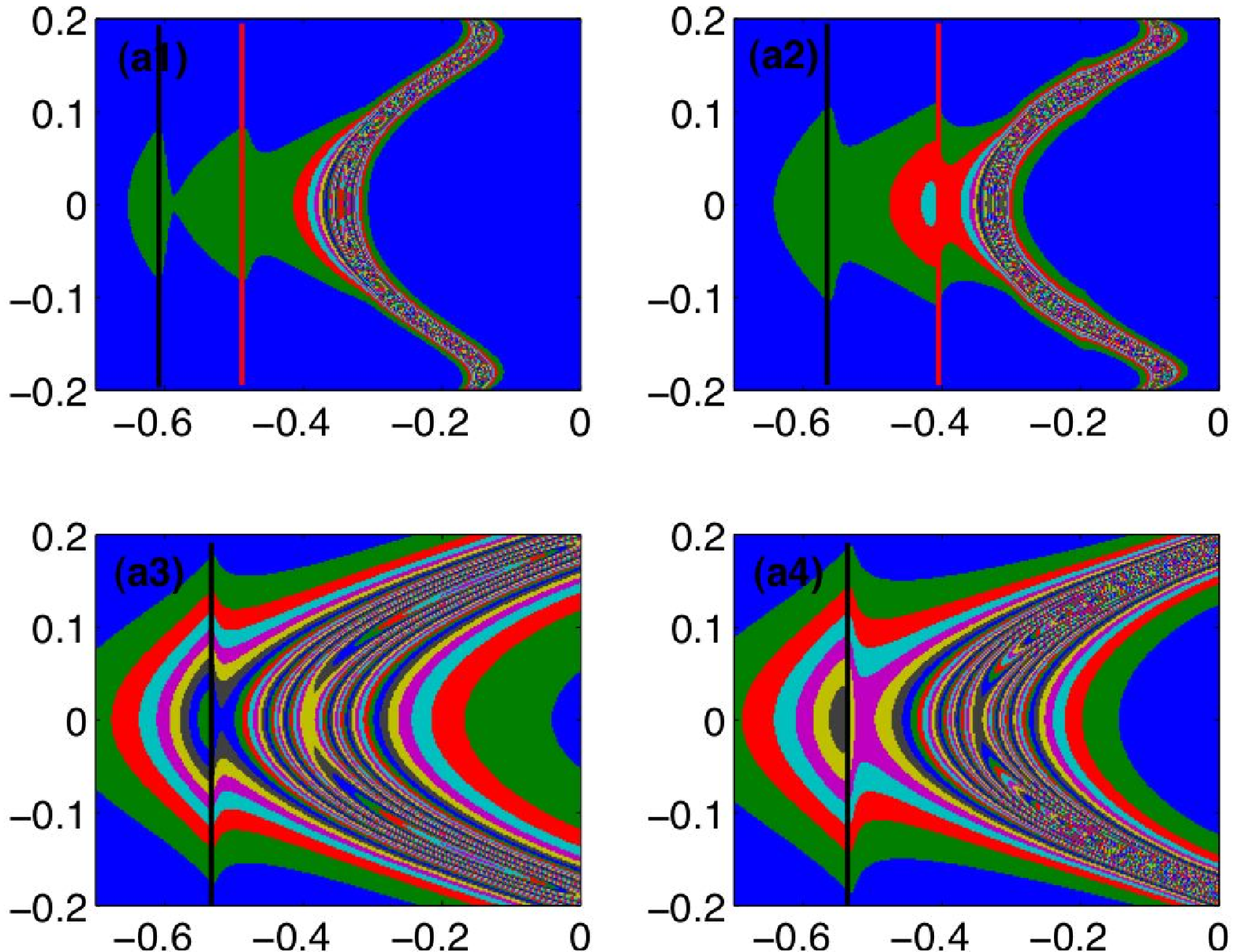,width=8.0cm,angle=0}}
\centerline{\psfig{figure=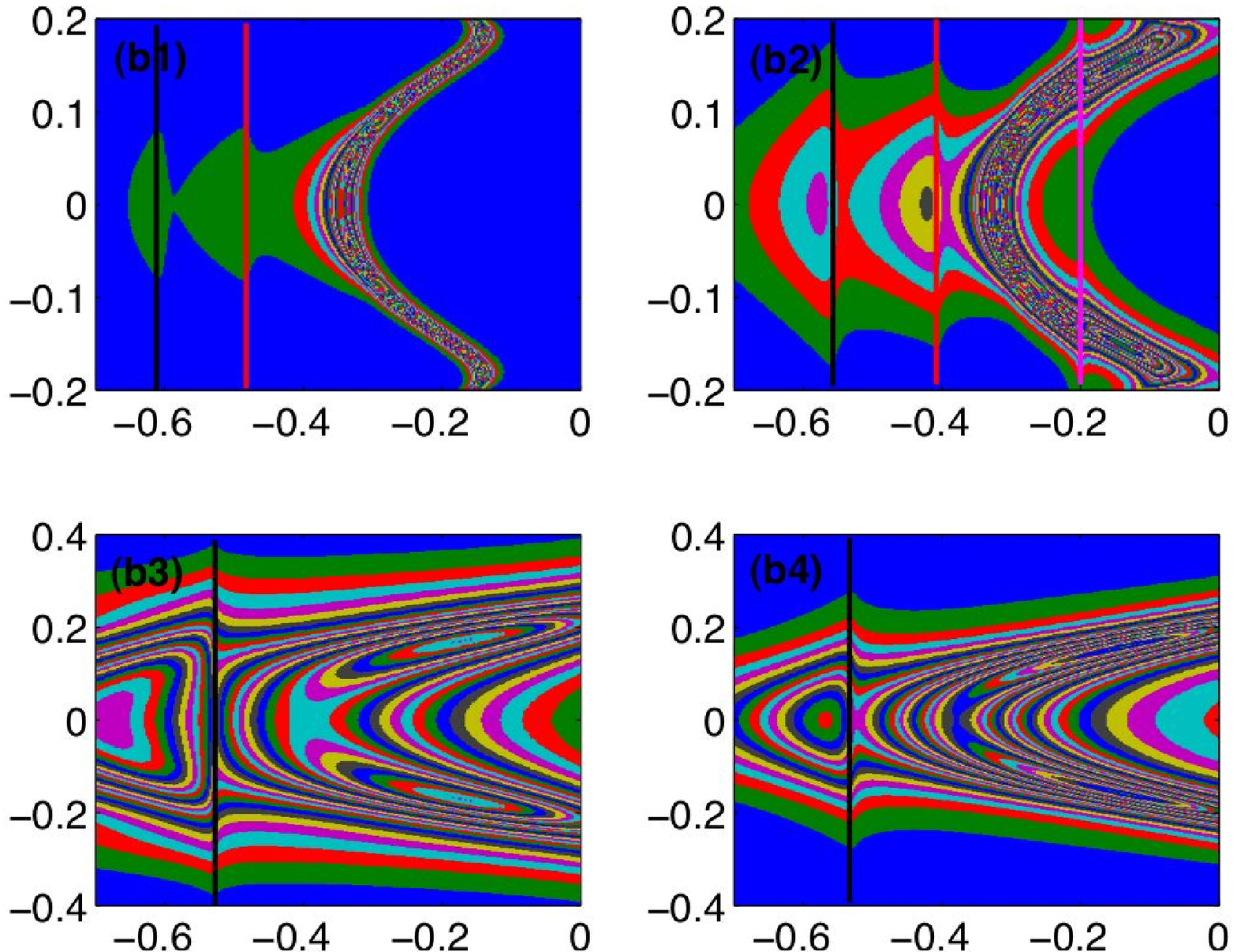,width=8.0cm,angle=0}}
\caption{(Color) Time-resolved spectral function for selectively
pumping electrons (a$_1$-a$_4$ for $\tau= -100, -2.7, 2.8, 600$) and phonons (b$_1$-b$_4$ for $\tau=-1000, -158, 110, 800$), respectively. The vertical axis is $k_x/\pi$ at a fixed $k_y = 0.75\pi$ and the horizontal axis is energy. The black, red, and magenta lines represent the energy location, 
$-(E_M+\Omega_0)$, $-(\Delta_0(T_e)+\Omega_0)$, and $-(\Omega_0 - \Delta_0(T_e))$, corresponding to the dip or hump location discussed in Fig.~\ref{dosf}.
}
\label{spctph}
\end{figure}
For direct pumping of phonons,
we choose different time sequences $\tau = -1000$, $-158$,
$110$, and $800$ to evaluate the DOS.   
The results are displayed in Fig.~\ref{dosf}(b). 
From Fig.~\ref{dosf}(b), we observe that the dips suggesting the el-ph couplig
in both normal and supercondcting states becomes stronger than those for 
selectively pumping electrons, as expected. In particular, we observe two
new small dips at $\omega=\pm 0.05$ for $\tau=-158$ (at which $T_e=0.041$ and
$T_{ph}=0.163$), both of which arise from the poles
of $\Sigma_\kv(\omega)$ at $\omega=\pm (\Omega_{0}-\Delta_0(T_e))$, 
in addition to those observed in the case of electrons being excited directly. 
This is because the hot phononic temperature becomes very high while the 
electronic temperature is still cold  and 
the contributions from the terms as weighted by $N(\pm \Omega_0)$ are significantly enhanced.
Figure.~\ref{dosf}(b) clearly shows that, as time elapses from 
$\tau=-1000$, $-158$, to $110$, 
the peaks move toward the {\em zero energy}, 
while from $\tau = 110$ to $800$,
the locations of these two peaks remain unchanged. 
This can be understood by considering that, after pumping, $T_e$ rises with the
time delay, while $\Delta_0(T_e)$ decreases and rapidly vanishes 
when $T_e$ reaches $T_c$. 

Finally we numerically evaluate the time-resolved spectral function for 
the two excitations discussed above at those time sequences chosen for 
calculating the DOS. 
In our calculations, we consider cuts along $k_x$-axis in the Brillouin zone 
at a $k_y$ chosen near the zone boundary.
Fig.~\ref{spctph} shows the snapshots of the image of the spectral 
function $A(\mathbf{k},\omega)$ 
as a function of $k_x$ and $\omega$.
From Fig.~\ref{spctph}(a$_1$,a$_2$), 
we observe kinks at $\omega=-(\Delta_0(T_e)+\Omega_0)$ (red lines) 
and $-(E_M+\Omega_0)$ (black lines)  at $\tau = -100$ and $-2.7$. The red line 
is shifted in energy as time 
elapses from  $\tau=-100$ to $\tau=-2.7$ because the BCS gap $\Delta_0 (T_e)$ 
drops significantly from 0.19 to 0.1. These kinks correspond to the dip 
structures discussed in Fig.~\ref{dosf}. As  $\tau > -2.3$, the superconducting 
gap vanishes, so do the kinks marked with red lines 
(see Fig.~\ref{spctph}(a$_3$-a$_4$)). The kinks marked with black lines 
persist through the whole time sequence, which can be ascribed to the el-ph 
coupling signature in the normal state. 
Therefore we have every reason to regard these kinks marked with red lines 
as the signature of the
el-ph coupling in the superconducting state. 
For hot phonon pumping (Fig.~\ref{spctph}(b$_1$-b$_4$)), 
in addition to those signatures of the el-ph coupling observed 
for electron pumping, 
new kink structure at $-(\Omega_0-\Delta_0(T_e))$(marked with magenta line in 
Fig.~\ref{spctph}(b$_2$) is also observed. These observations
are consistent with our finding in the DOS. 
In the viewpoint that the electronic spin fluctuations arise from the 
strong correlation within the electronic DoF itself, and if one can 
assume that the spin fluctuations will ride on the electrons and thus 
its effective temperature will be tied to the electronic temperature, 
direct pumping of hot phonons will provide a unique way to 
differentiating the bosonic modes being of electronic or phononic origin, 
to which electronic quasiparticles are strongly coupled in HTSC.  

In conclusion, we have developed a three-temperature model to simulate the time
evolution of temperature for electrons, hot phonons, and the lattice. Our model
takes both quasiparticle excitation and relaxation into account and thus
is valid in both normal and superconducting states, because at the early stage
of excitation, the system can be still in the supercondcting state.
Based on this model, we derive a formula for the time-resolved spectral 
function, allowing us to investigate the dynamics of the el-ph coupling for 
HTSC materials. 
 
{\bf Acknowledgments.} We thank A. V. Balatsky, Elbert E. M. Chia, 
Hari Dahal, J. K. Freericks, M. Graf, A. Piryatinski, A. J. Taylor, 
S. A. Trugman, and D. Yarotski for 
valuable discussions. This work was supported by the National Nuclear 
Security Administration of the U.S. DOE  at  LANL under Contract 
No. DE-AC52-06NA25396, the U.S. DOE 
Office of Science, and the LDRD Program at LANL.

\end{document}